\documentclass[%
 reprint,
superscriptaddress,
 amsmath,amssymb,
 aps,
floatfix,
]{revtex4-2}

\usepackage{graphicx}% Include figure files
\usepackage{dcolumn}% Align table columns on decimal point
\usepackage{bm}% bold math
\usepackage{siunitx}
\begin{document}
 
\preprint{APS/123-QED}

\title{Characterization of a Disordered Above Room Temperature Skyrmion Material Co$_{8}$Zn$_{8}$Mn$_{4}$}% Force line breaks with \\

\author{M. E. Henderson}
\email{mehenderson@uwaterloo.ca}%
\affiliation{Institute for Quantum Computing, University of Waterloo, Waterloo, ON, Canada, N2L3G1}
\affiliation{Department of Physics \& Astronomy, University of Waterloo,
  Waterloo, ON, Canada, N2L3G1}

\author{J. Beare}
\affiliation{Department of Physics and Astronomy, McMaster University, Hamilton, ON, Canada, L8S 4M1}
\author{S. Sharma}
\affiliation{Department of Physics and Astronomy, McMaster University, Hamilton, ON, Canada, L8S 4M1}
\author{M. Bleuel}
\affiliation{National Institute of Standards and Technology, Gaithersburg, Maryland 20899, USA}
\affiliation{Department of Materials Science and Engineering, University of Maryland, College Park, MD  20742-2115}
\author{P. Clancy}
\affiliation{Brockhouse Institute for Materials Research, Hamilton, ON, Canada, L8S 4M1}
\author{D. G. Cory}
\affiliation{Institute for Quantum Computing, University of Waterloo, Waterloo, ON, Canada, N2L3G1}
\affiliation{Department of Chemistry, University of Waterloo, Waterloo, ON, Canada, N2L3G1}
\author{M. G. Huber}
\affiliation{National Institute of Standards and Technology, Gaithersburg, MD 20899, USA}
\author{C. A. Marjerrison}
\affiliation{Brockhouse Institute for Materials Research, Hamilton, ON, Canada, L8S 4M1}
\author{M. Pula}
\affiliation{Department of Physics and Astronomy, McMaster University, Hamilton, ON, Canada, L8S 4M1}
\author{D. Sarenac}
\affiliation{Institute for Quantum Computing, University of Waterloo, Waterloo, ON, Canada, N2L3G1}
\author{E. M. Smith}
\affiliation{Department of Physics and Astronomy, McMaster University, Hamilton, ON, Canada, L8S 4M1}
\author{K. Zhernenkov}
\affiliation{Institute for Quantum Computing, University of Waterloo, Waterloo, ON, Canada, N2L3G1}
\affiliation{J\"ulich Centre for Neutron Science at Heinz Maier-Leibnitz Zentrum, Forschungszentrum J\"ulich GmbH, 85748 Garching, Germany}
\author{G. M. Luke}
\affiliation{Department of Physics and Astronomy, McMaster University, Hamilton, ON, Canada, L8S 4M1}
\affiliation{Brockhouse Institute for Materials Research, Hamilton, ON, Canada, L8S 4M1}
\author{D. A. Pushin}
\email{dmitry.pushin@uwaterloo.ca}%
\affiliation{Institute for Quantum Computing, University of Waterloo, Waterloo, ON, Canada, N2L3G1}
\affiliation{Department of Physics \& Astronomy, University of Waterloo,
  Waterloo, ON, Canada, N2L3G1}

\date{\today}

\begin{abstract}
Topologically non trivial spin textures host great promise for future spintronic applications. Skyrmions in particular are of burgeoning interest owing to their nanometric size, topological protection, and high mobility via ultra-low current densities. It has been previously reported through magnetic susceptibility, microscopy, and scattering techniques that Co$_{8}$Zn$_{8}$Mn$_{4}$ forms an above room temperature triangular skyrmion lattice. Here we report the synthesis procedure and characterization of a polycrystalline Co$_{8}$Zn$_{8}$Mn$_{4}$ bulk sample. We employ powder x-ray diffraction, backscatter Laue diffraction, and neutron diffraction as characterization tools of the crystallinity of the samples, while magnetic susceptibility and Small Angle Neutron Scattering (SANS) measurements are performed to study the skyrmion phase. Magnetic susceptibility measurements show a dip anomaly in the magnetization curves which persists over a range of approximately 305 K- 315 K. SANS measurements reveal a rotationally disordered polydomain skymrion lattice. Applying a recently developed symmetry-breaking magnetic field sequence, we were able to orient and order the previously jammed state to yield the prototypical hexagonal diffraction patterns, with secondary diffraction rings.\end{abstract}

\maketitle

%\tableofcontents

\section{\label{sec:level1}Introduction\protect\\}

Originally proposed by Tony Skyrme to explain the stability of hadrons \cite{Skyrme1962unified}, skyrmions have since emerged as promising topological objects in condensed matter systems ranging from Bose-Einstein condensates \cite{khawaja2001skyrmions,Leslie2009creation} to quantum Hall systems \cite{sondhi1993skyrmions,makeover}. Notably, magnetic manifestations of such objects have been experimentally realized in noncentrosymmetric materials, stabilized by the antisymmetric Dzyaloshinskii–Moriya (DM) exchange interaction \cite{muhlbauer2009skyrmion,Yu2011near,Munzer2010skyrmion,Shibata2013towards}. These so called magnetic skyrmions exist as localized, nanometric sized, topologically stable, spin vortices characterized by a quantized topological charge \cite{makeover}. This notion of topological charge forbids a continuous change of homotopy, mathematically speaking. In real physical systems, this translates to an energy landscape for transitions between topologically distinct states, invoking a sense of robustness against deformations into topologically trivial spin textures (i.e. those possessing a topological charge of zero) \cite{Hagemeister2015stability,schutte2014dynamics,Bessarab2015method}. Their nontrivial topological spin textures, and resultant emergent electromagnetic fields \cite{Schulz2012emergent}, give rise to novel phenomena such as a topological Hall effect \cite{Kanazawa2011large,Neubauer2009topological}, multiferroic behavior \cite{Seki2012observation,White2012Electric,Ruff2015multiferroicity}, and current driven dynamics five to six orders of magnitude smaller than those currently required to drive domain walls in ferromagnets \cite{Jonietz2010spin,Schulz2012emergent,soccer}. Ultimately, their electric controllability, in combination with their nanometric size, make magnetic skyrmions prime candidates for potential information carriers in quantum information science \cite{Zhang2015magnetic,Wiesendanger2016nanoscale, Yu2012skyrmion,Iwasaki2013current,Sampaio2013nucleation}.

Magnetic skyrmions are known to occur in materials lacking inversion symmetry, owing to chiral crystal structures in bulk magnets \cite{muhlbauer2009skyrmion,Wilhelm2011precursor,Munzer2010skyrmion,Seki2012observation,roomtemp,Kezsmarki2015neel} and thin films \cite{Tonomura2012Real,Yu2011near,Yu2010real,Nagase2019smectic}, or non-equivalent interfaces in multilayers and ultra-thin films \cite{Heinze2011spontaneous,Romming2013writing,Dupe2014tailoring,thinfilm}. In noncentrosymmetric chiral lattices the competition of the Heisenberg exchange interaction with the antisymmetric DM exchange interaction \cite{Dzyaloshinskii1958thermodynamic,Moriya1960anisotropic} tends to stabilize helical ground states in ferromagnetic crystals \cite{multi}. The application of a laboratory magnetic field breaks the symmetry of the helical ground state, and the superposition of three helical waves in the plane perpendicular to the laboratory field generates a two-dimensional triangular lattice of skyrmions. Noncentrosymmetric helimagnets in B20-type alloys such as MnSi \cite{muhlbauer2009skyrmion,Tonomura2012Real}, $\mathrm{Fe}_{1-x}\mathrm{Co}_{x}\mathrm{Si}$ \cite{Munzer2010skyrmion} and FeGe \cite{Yu2011near,FeGe} (all of which possess the same cubic chiral space group) have been shown to support sub ambient temperature skyrmion phases \cite{roomtemp}. However, skyrmion formation below room temperature presents an inherent implementation challenge for spintronic applications. 
Ref. \cite{roomtemp} reported $\beta-\mathrm{Mn}$ type Co-Zn-Mn alloys, specifically Co$_{8}$Zn$_{8}$Mn$_{4}$, to be triangular lattice room temperature skyrmions via Small Angle Neutron Scattering (SANS), Lorentz Transmisson Electron Microscopy (LTEM), and magnetization measurements \cite{roomtemp}. Here we provide a detailed synthesis and characterization procedure of a bulk polycrystalline Co$_{8}$Zn$_{8}$Mn$_{4}$ samples under both argon and air atmospheres. We report the application of a technique developed in \cite{dustin} to precipitate ordered and oriented skyrmions, which yields secondary diffraction rings.    

\section{\label{sec:level2}Synthesis\protect\\}
The material was synthesized via the solid state reaction 8Co + 8Zn + 4Mn $\rightarrow$ $\mathrm{Co}_{8}\mathrm{Zn}_{8}\mathrm{Mn}_{4}$. The powders were mixed in stoichiometric ratios in an agate mortar under an argon atmosphere. Once thoroughly ground, the resulting mixture was pressed into a pellet, which was then sealed in an evacuated quartz tube with a conically shaped end. The conical shape of the ampoule served to facilitate nucleation along a dominant growth direction, imposed by the geometry of the confining tube. The ampoule was inserted into a furnace at $700^{\circ}\mathrm{C}$ and the temperature was increased to $1025^{\circ}\mathrm{C}$ over the course of 12 hours. It was then cooled at a rate of $2^{\circ}\mathrm{C/{h}}$ until $900^{\circ}\mathrm{C}$ was reached. Finally,  it was cooled to $700^{\circ}\mathrm{C}$ over four hours and was removed. The final product was a conical shaped silver polycrystal (approximately 2-3 grains) with dimensions 0.8 cm x 1.4 cm (diameter x length) and mass of 2 g as shown in Fig. 1a).
\begin{figure}[b]\centering\includegraphics[width = \columnwidth]{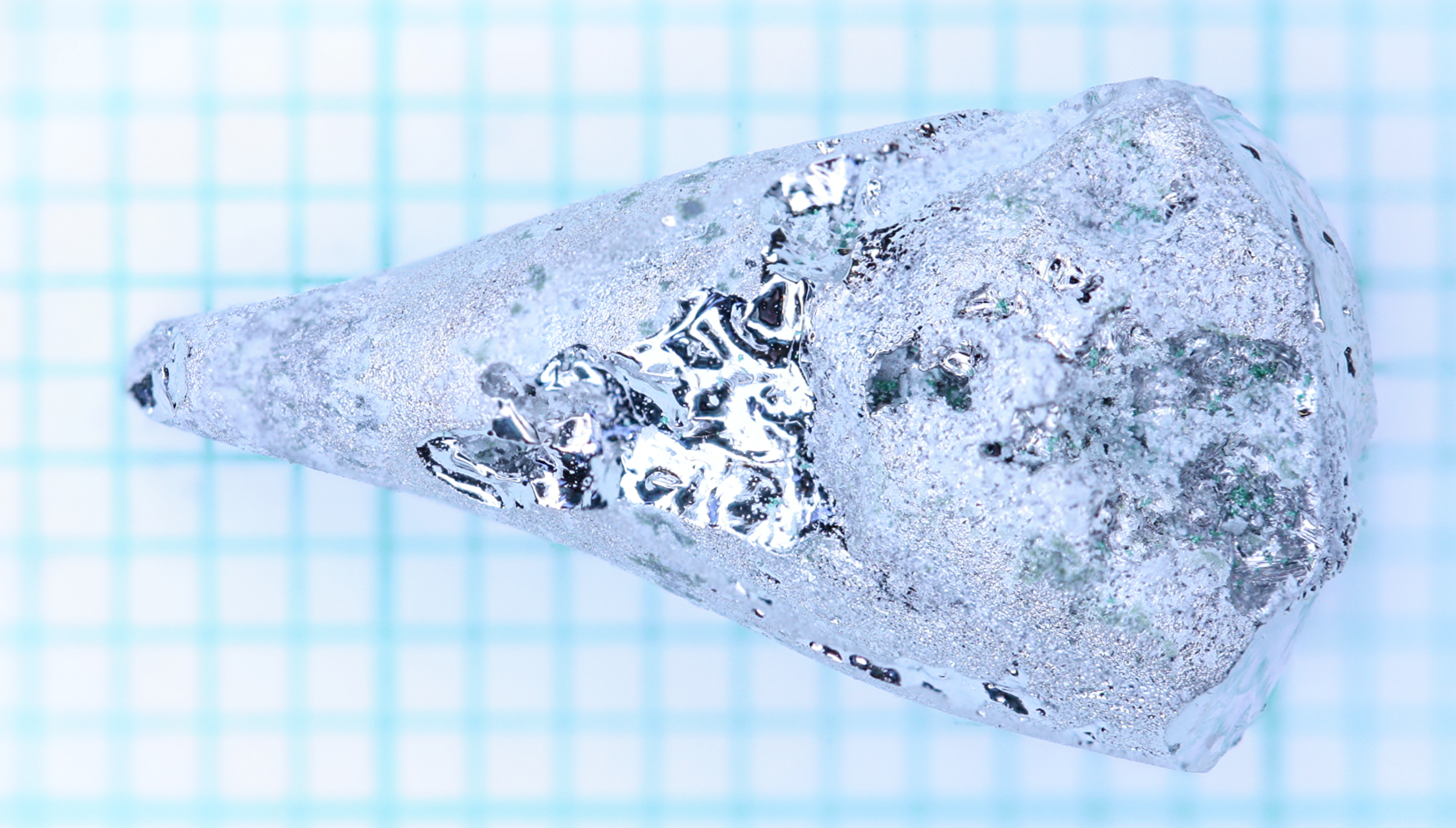}
\caption{\label{fig:epsart1}Polycrystalline Co$_{8}$Zn$_{8}$Mn$_{4}$ sample mixed under argon of dimensions 0.8 cm x 1.4 cm and mass 2 g. Each grid line corresponds to 1 mm.}
\end{figure}

\begin{figure}\includegraphics[width = \columnwidth]{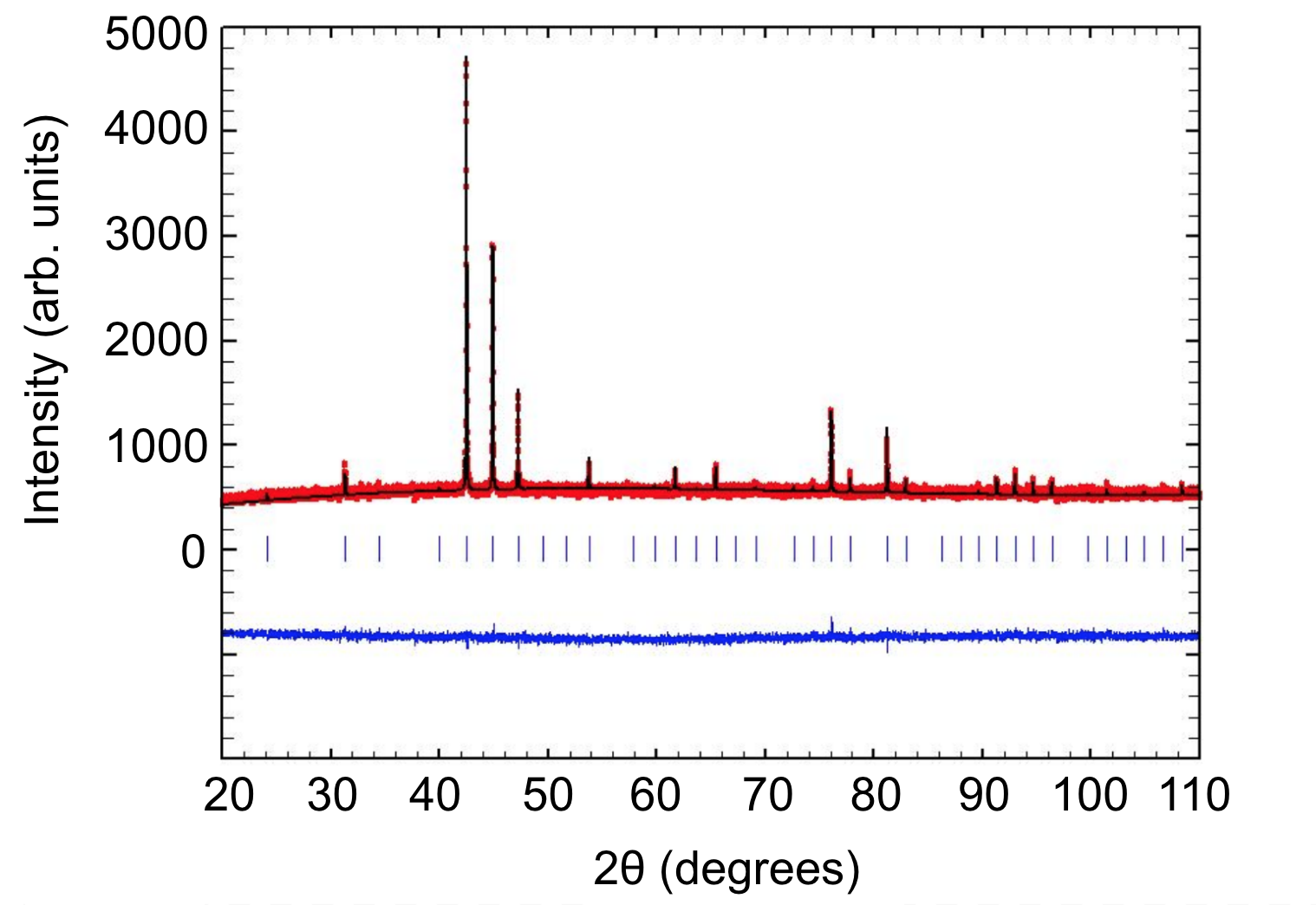}
\caption{\label{fig:powderXRD} Rietveld refinement for powder X-ray diffraction of Co$_{8}$Zn$_{8}$Mn$_{4}$. The black curve is the predicted spectra, the red line is the data, and the blue is the difference between the two. The blue vertical lines indicate the locations of the expected peaks. The refinement demonstrates the sample is phase pure with space-group $ \beta $-Mn and lattice constant $6.37161(1)\si{\angstrom}$. }
\end{figure}

The reaction products were analysed via powder X-ray diffraction in the scattering angular ($2\theta$) range of $20^{\circ}-110^{\circ}$ using the Cu $K_{\alpha 1}$  wavelength of $1.5406\si{\angstrom}$. A Rietveld refinement of the diffraction data to the $P4_{1}32$ space-group ($ \beta $-Mn-type) was performed using the FullProf program, from which we were able to extract a lattice constant of $6.37161(1)\si{\angstrom}$. Fig. \ref{fig:powderXRD} shows the Rietveld refinement for the powder x-ray diffraction spectra, where the red dots are the measured spectra, the black line is the predicted spectra (where the vertical blue lines below indicate expected peak locations), and the blue line is the difference between the two.  The sharpness of present peaks (evidenced by the zero slope of the blue curve), and absence of additional peaks indicate the sample is phase pure.       

\section{\label{sec:level3}Characterization\protect\\}
    
\begin{figure}[ht]\includegraphics[scale =0.5]{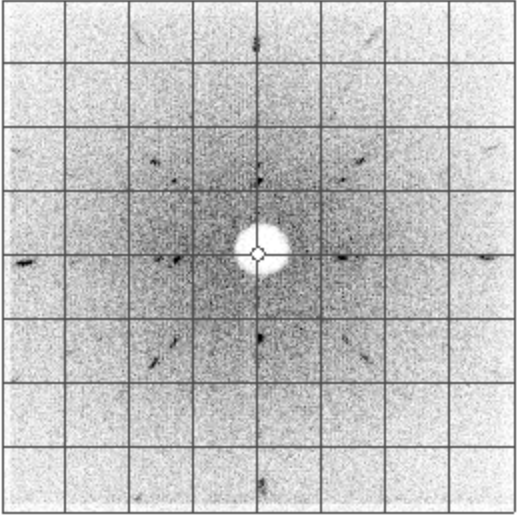}
\caption{\label{fig:Laue} Laue image along (100) direction of dominant grain on the front face of the cube. All peaks were indexed to the (100) direction, verifying the single grain portion of this material. The fourfold symmetry of the pattern is characteristic for the (100) direction of a cubic crystal.}
\end{figure}

Backscatter X-ray Laue diffraction was performed as a preliminary investigation of the crystallinity and orientation of the material (Fig. \ref{fig:Laue}). Based on a changing diffraction pattern during translation, we were able to identify grain boundaries.  Through systematic scanning and slicing, the polycrystal was cut into a rectangular prism of dimensions 3.4 mm x 3.3 mm x 3.0 mm while mapping the crystal orientation of the polycrystalline sample. The final product was polycrystalline with the (100) direction of the dominant grain along one face of the rectangular prism. Fig.  \ref{fig:Laue} shows a Laue pattern for the dominant grain, demonstrating the archetypal 4-fold symmetry of the cubic lattice along the (100) direction.  

\begin{figure}[ht]\includegraphics[width = \columnwidth]{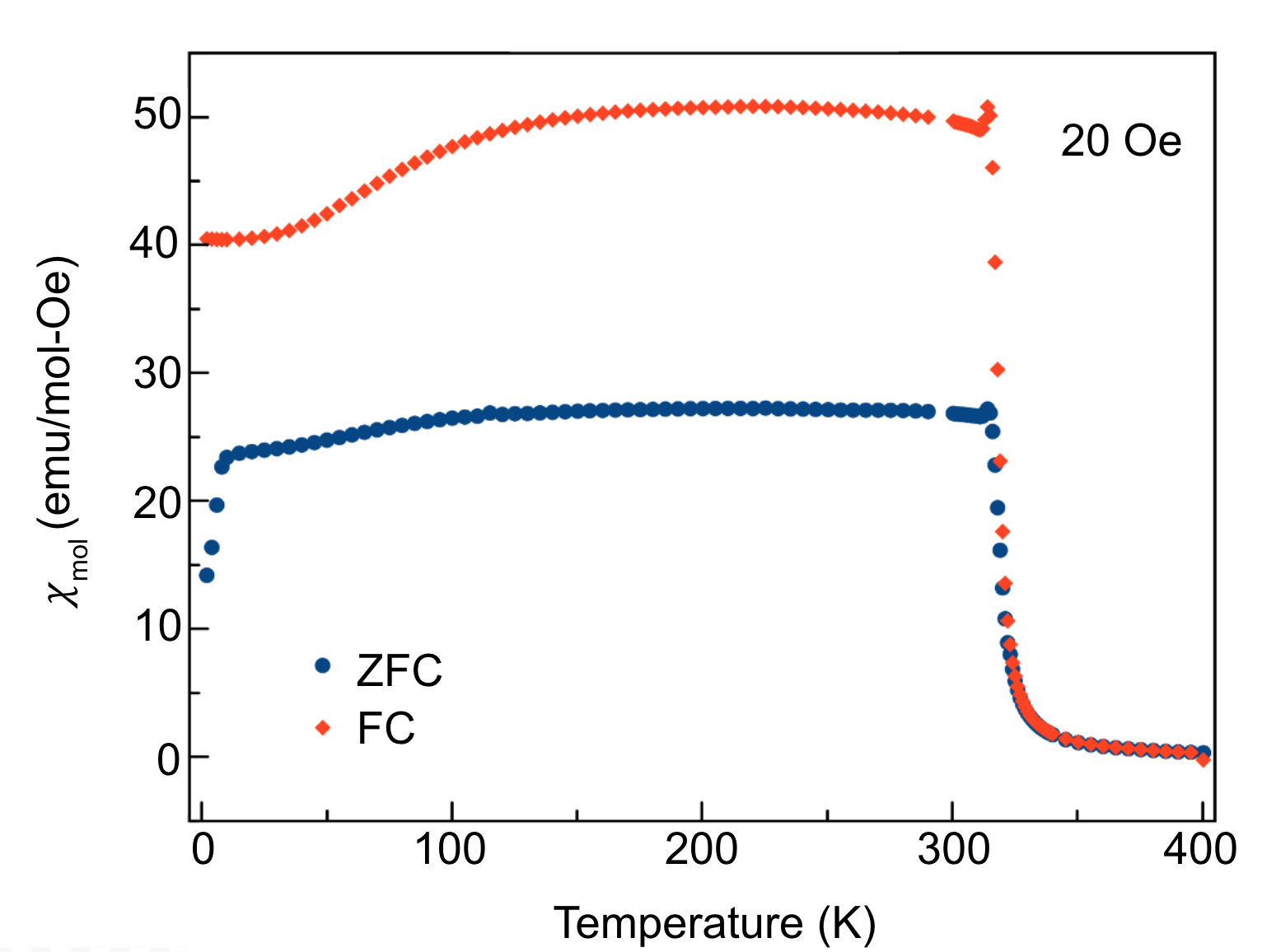}
\caption{\label{fig:DC_Susceptibility} Magnetic susceptibility per mol of Co$_{8}$Zn$_{8}$Mn$_{4}$ after zero field cooling (ZFC) and field cooling from 400 K in a magnetic field of 20 Oe. The bifurcation behaviour of the ZFC and FC curves at 7 K results from the path-dependent behaviour of the susceptibility, which is indicative of a spin glass transition. }
\end{figure}

Magnetic susceptibility measurements were performed using a Quantum Design MPMS 5 Superconducting Quantum Interference Device (SQUID) with an AC option installed. The high temperature ferromagnetic phase was verified via field cooling (FC) from 400 K (Fig. \ref{fig:DC_Susceptibility}).
The onset of the transition was found to be 320 K, consistent with Ref. \cite{roomtemp}. A Curie-Weiss fit between 350 K and 400 K results in an effective magnetic moment of 1.6 $\mu_{B}$, consistent with the magnetic moment found for other Co$_{x}$Zn$_{y}$Mn$_{z}$ (x+y+z=20) compounds \cite{Bocarsly2019}. The high temperature hysterisis between the FC and ZFC (zero field cooled) curves (vertical offset) is a result of the disparity in magnetization due to the aligned ferromagnetically ordered domains in the FC case as opposed to the misaligned domains in the ZFC case which produce a smaller commensurate moment.  Upon further cooling there was a notable path dependence, as is evident by the sharp change in temperature dependence in the ZFC magnetization at around 7 K in Fig. \ref{fig:DC_Susceptibility}. \begin{figure}[ht]\includegraphics[width = \columnwidth]{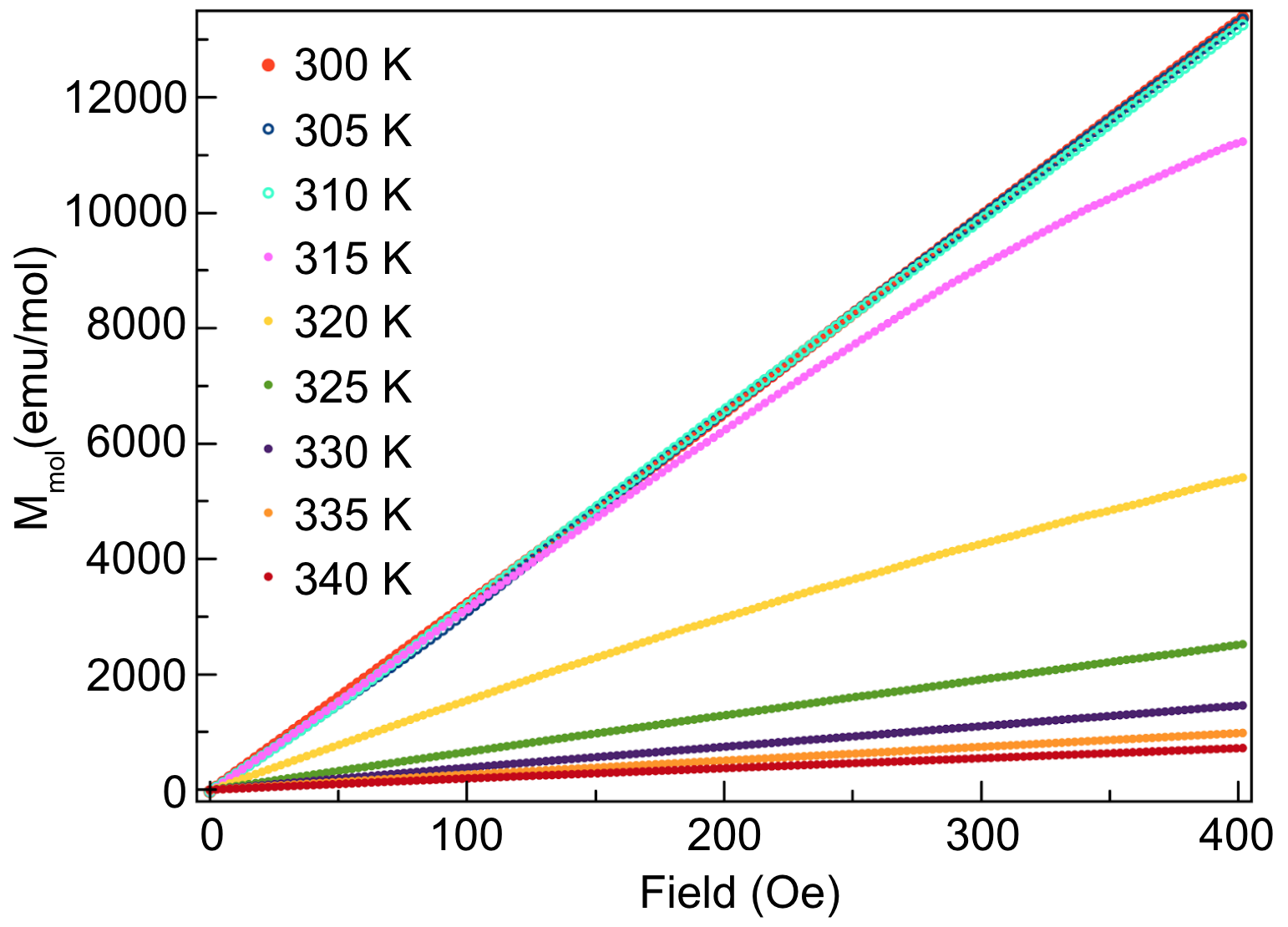}
\caption{\label{fig:MvsH} Magnetic field dependent magnetization upon increasing magnetic fields from 0 Oe to 400 Oe for a temperature range of 300 K-340 K, in 5 K increments. Note a substantial decrease in magnetization for temperatures greater then 320 K, consistent with a paramagnetic phase.}
\end{figure} This magnetic hysteresis is evidence of a spin glass transition \cite{Nagata1979}. The mechanisms underpinning this transition have been previously reported via crystal structure analysis by neutron powder diffraction in \cite{spin} as occurring due to site mixing between the Co and Mn atoms on the 8c crystallographic sites, which gives rise to random competition amongst the ferromagnetic and antiferromagnetic interactions, yielding quenched magnetic disorder.
Further susceptibility measurements were carried out to confirm the presence of the skyrmion phase; Fig. \ref{fig:MvsH} shows isothermal magnetization measurements as a function of magnetic field  for a 20.0 mg polycrystalline piece of the sample. \begin{figure}[t]\includegraphics[width = \columnwidth]{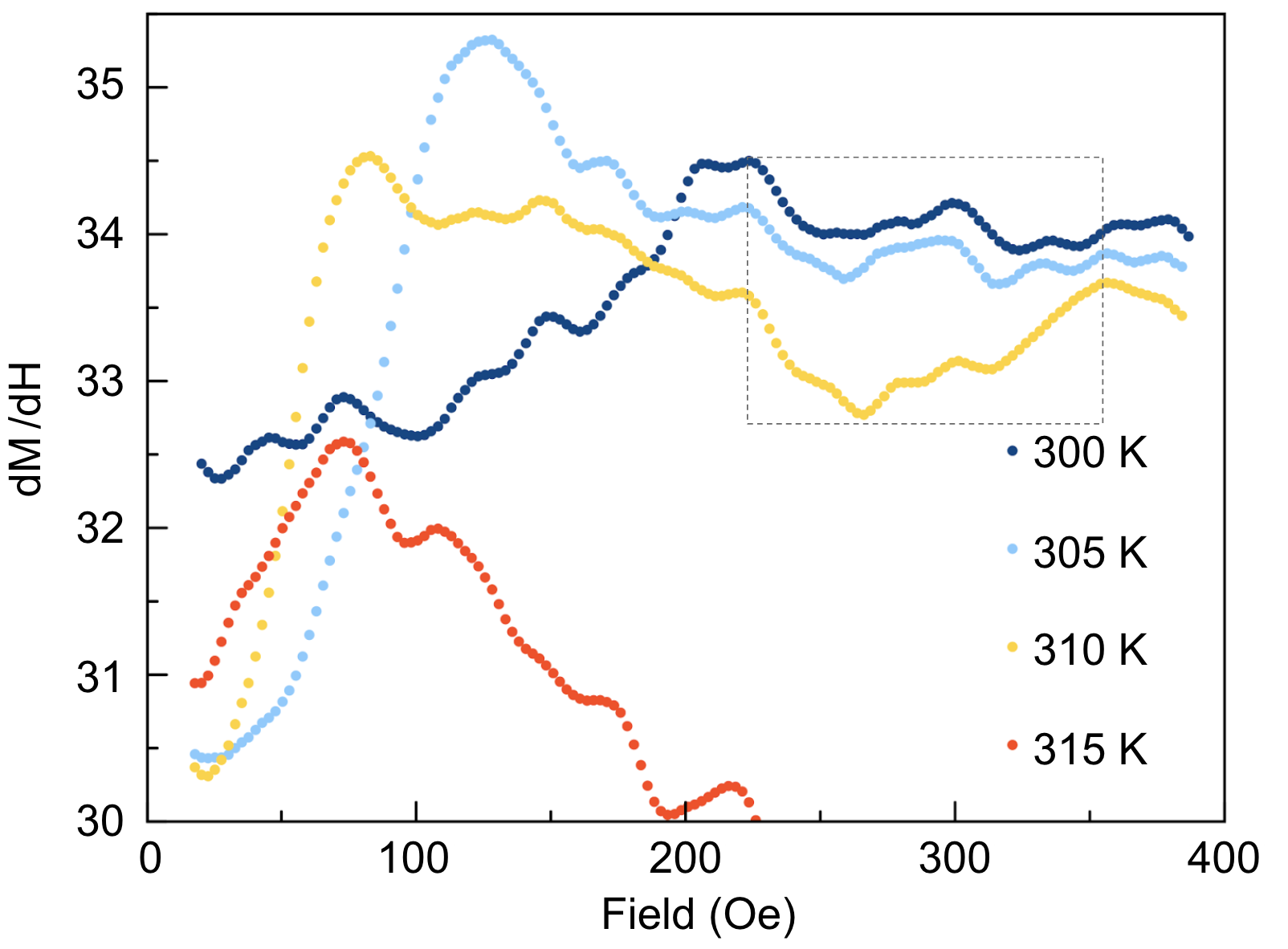}
\caption{\label{fig:XvsH} Temperature dependent isothermal differential magnetic susceptibility upon increasing magnetic fields from 0 Oe to 400 Oe. The dip structure (region contained within the rectangular dotted box) is most clearly pronounced for 310 K, and presents at a field of $\sim200$ Oe, indicating the onset of the skyrmion phase.}
\end{figure}The magnetization measurements were taken while increasing the DC field from 0 Oe to 400 Oe, after which measurements were taken while decreasing the field (not shown). The notable decrease in magnetization for temperatures above 320 K is consistent with exiting an ordered phase into a paramagnetic phase.
\par
\begin{figure}[ht]
\includegraphics[width = \columnwidth]{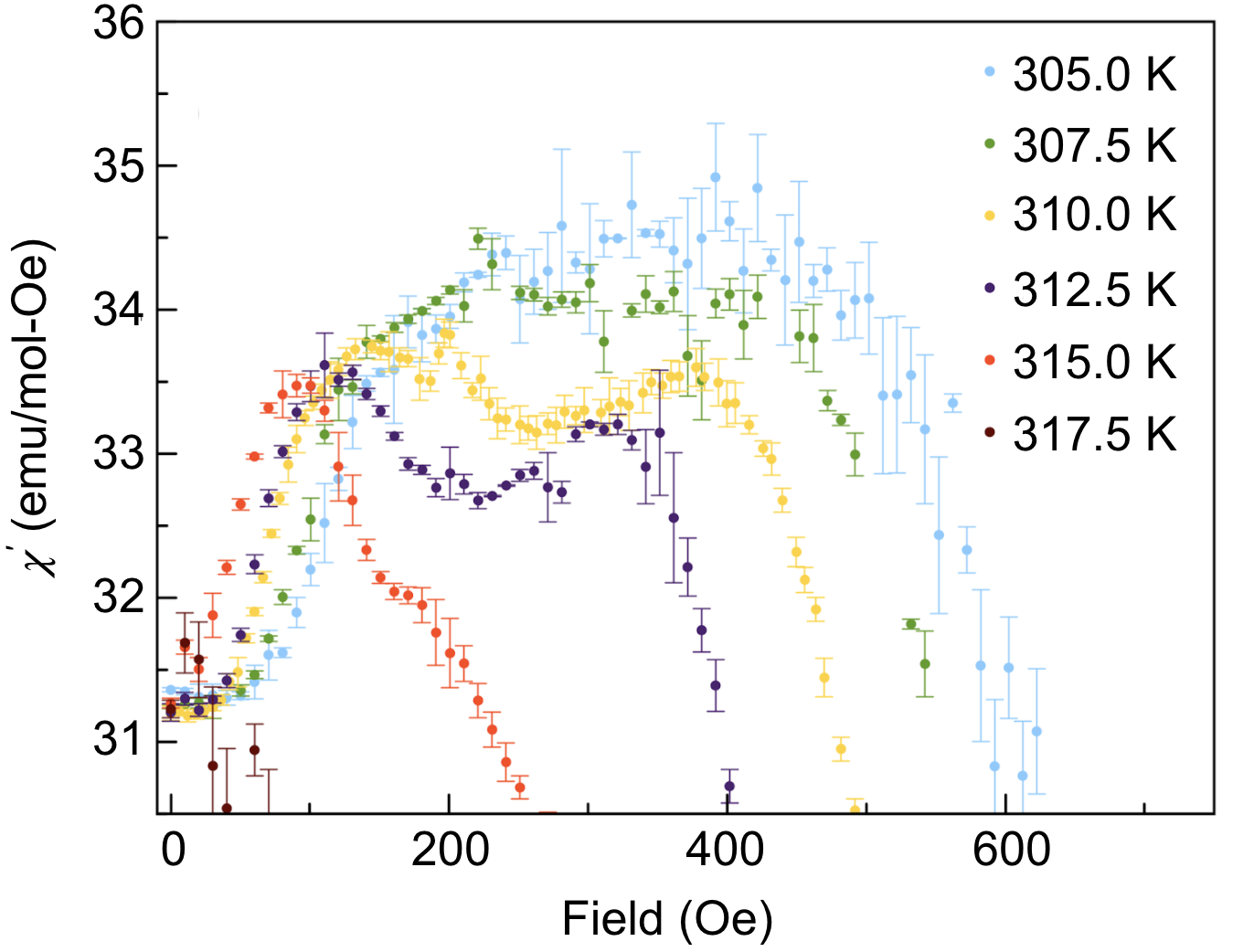}
\caption{\label{fig:realX} Temperature dependence of AC magnetic susceptibility per mol of Co$_{8}$Zn$_{8}$Mn$_{4}$ over a range of 305 K-317.5 K after increasing magnetic fields from 0 Oe to 500 Oe in a 100 Hz driving field, with an amplitude of 0.1 Oe. The skyrmion phase is most pronounced at a temperature of 310 K, and is observed to persist in the dip anomaly between 100 Oe and 450 Oe. The field value at the minimum of the dip determines the largest and most robust skyrmion phase; these temperature and field parameters are then used for SANS measurements on the material.}
\end{figure}

 We performed differential magnetic susceptibility measurements at 300 K, 305 K, and 310 K. This was investigated by taking numerical derivatives of the M vs H curves and, after smoothing the data,  show abrupt dips in the susceptibility (with the strongest dip occurring at 310 K shown by the yellow curve in the rectangular dotted box in Fig. \ref{fig:XvsH}), suggestive of a phase transition consistent with \cite{roomtemp}.
AC susceptibility measurements were performed which depend upon $\frac{dM}{dH}$ but do not involve using a numerical derivative, which can be susceptible to large fluctuations. AC measurements are therefore a much more sensitive technique, yielding a much smaller uncertainty than the above differential magnetic susceptibility measurements. The AC susceptibility measurements show similar peaks as a function of applied field, indicating a phase transition. The most pronounced dip structure is again observed at a temperature of 310 K (yellow curve in Fig. \ref{fig:realX}), consistent with Fig. \ref{fig:XvsH}. These dip structures are well-known markers of the temperatures and fields over which the skyrmion phase exists \cite{roomtemp}. The AC susceptibility shows much cleaner and more defined dip structures than \ref{fig:XvsH}. AC measurements also probe the dynamics of the system and may be included in future work to probe the time scales of the metastable skyrmion phases found below 300 K.

We performed unpolarized SANS at the NG7-30m beamline at the National Institute for Standards and Technology (NIST) for a 15 m beam configuration and a neutron wavelength of $6\si{\angstrom}$ \cite{30mGlinka2018,kline2006reduction,disclaimer}. At room temperature in zero field, our initial SANS measurements revealed four smeared magnetic satellites atop a circular ring, indicating multi-domain single q-helical structures, with the preferential smearing direction of the peaks elucidating the anisotropy direction (Fig. 8a). Upon field cooling through the ferromagnetic phase from 420 K to 310 K, in a field of 250 Oe, a ring developed. The absence of the signature triangular lattice skyrmion hexagonal pattern is a result of the polycrystalline nature of the material; the misalignment of the skyrmion domains breaks the order in many directions thereby smearing the hexagonal patterns, precipitated by each individual domain, to produce a ring. \begin{figure}
\includegraphics[width = \columnwidth]{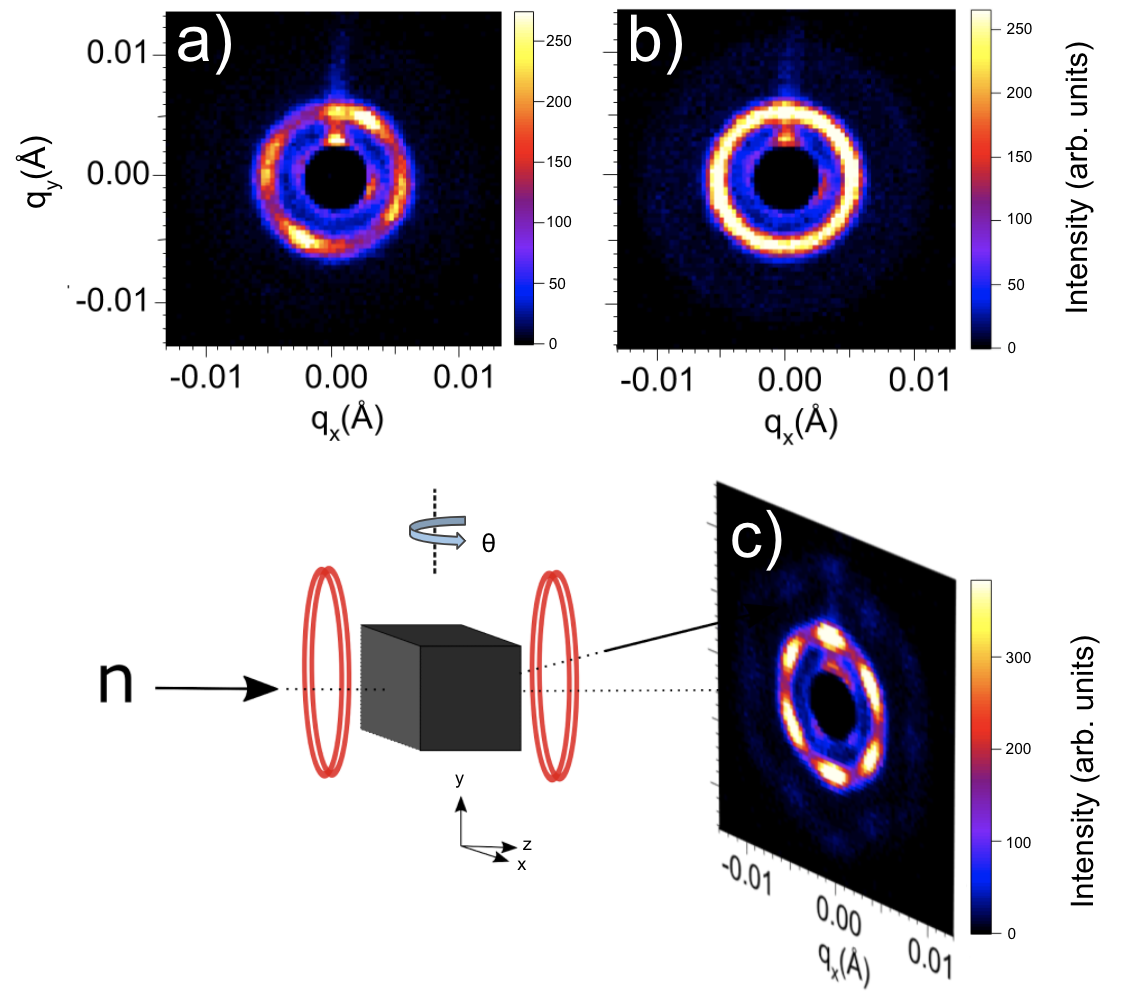}
\caption{\label{fig:epsart2} SANS images showing disordered helical ground state at room temperature in zero field a), initial scattering ring for disordered skyrmion domains at 310 K in a magnetic field of 250 Oe b), schematic of symmetry-breaking field rotation setup, and SANS image after 30 rotations of the rotation sequence at 310 K in a field of 250 Oe, c). The count times of the images are 900s, 711s, and 600s, respectively. The increased intensity/preferential smearing of the peaks in the top right and bottom left diagonals of the fourfold helical image elucidate the anisotropy direction for the crystal. Schematic of the rotation setup illustrates neutron propagation direction (n) is in the z direction. For the symmetry-breaking rotation sequence the sample is rotated symmetrically in the xy scattering plane about $\theta$, with the magnetic field fixed in the z direction.   }
\end{figure}Using a symmetry-breaking magnetic field sequence \cite{dustin} where the sample is rotated symmetrically in the static magnetic field to precipitate ordered and oriented skyrmion lattices despite the overwhelming structural disorder, the underlying triangular lattice skyrmion phase was revealed. The development of a first order ring with 6 peaks and an additional secondary ring was observed after 10 symmetric rotations. Fig. 8c) shows the fully discernible 6-fold primary ring after 30 rotations, accompanied by a second order ring mimicking the same hexagonal symmetry with 12 peaks. The presence of the secondary ring indicates potential multiple scattering and/or higher order diffraction. The underlying mechanism is left to be investigated for future experiments. 

In conclusion, we have successfully verified the synthesis and characterization of the above room temperature bulk triangular lattice skyrmion material Co$_{8}$Zn$_{8}$Mn$_{4}$ originally discovered in \cite{roomtemp}. Powder x-ray diffraction studies revealed a pure phase, while backscatter Laue diffraction and neutron diffraction indicated a polycrystalline material. SANS measurements demonstrated the underlying rotationally disordered skyrmion domains. Application of the symmetry-breaking rotation sequence developed in \cite{dustin} precipitated ordered and oriented triangular skyrmion lattices, yielding secondary diffraction rings. These secondary diffraction rings are most likely a combination of double scattering (owing to the thickness of the sample), and higher-order diffraction, in turn elucidating the effectiveness of the technique in \cite{dustin} for ordering and even promoting the growth of skyrmions, despite the presence of disorder, thereby producing long-range order. Future experiments may explore the ratio of multiple scattering to higher-order diffraction through the use of Renninger scans \cite{adams2011long}. Furthermore, we intend to use a newly developed reconstruction algorithm \cite{heacock2018sub} to perform 3D tomography of skyrmion topological transitions in the bulk, as well as incorporate spin components to explore the structure of the neutron wavefunction after passing through a skyrmion sample \cite{nsofini2016spin,Sarenac2019,Sarenac2018}.

\begin{acknowledgments}
This work was supported by the Canadian Excellence Research Chairs (CERC) program, the Natural Sciences and Engineering Council of Canada (NSERC) Discovery program, Collaborative Research and Training Experience (CREATE) program, the Canada First Research Excellence Fund (CFREF), and the National Institute of Standards and Technology (NIST) Quantum Information Program. Access to SANS and CHRNS was provided by the Center for High Resolution Neutron Scattering, a partnership between the National Institute of Standards and Technology and the National Science Foundation under Agreement No. DMR-1508249. We thank Jeff Krzywon for his help with the SANS setup, and Dustin Gilbert for his symmetry-breaking magnetic field sequence and numerous discussions.    
\end{acknowledgments}

\nocite{*}

\bibliography{refs}% Produces the bibliography via BibTeX.

\end{document}